\documentclass[superscriptaddress, reprint,longbibliography]{revtex4-1}
\usepackage{amsmath}
\usepackage{amssymb}
\usepackage{graphicx}
\usepackage{epstopdf}
\usepackage[utf8]{inputenc}
\usepackage[caption=false]{subfig}

\begin{document}
	
	\title{Pattern formation in a fully-3D segregating granular flow}
	\date{\today}
	
	\author{Mengqi Yu}
	\affiliation{Department of Chemical and Biological Engineering, Northwestern University, Evanston, IL 60208, USA}
	\author{Paul B. Umbanhowar}
	\affiliation{Department of Mechanical Engineering, Northwestern University, Evanston, IL 60208, USA}
	%\email[]{umbanhowar@northwestern.edu}
	\author{Julio M. Ottino}
	\affiliation{Department of Chemical and Biological Engineering, Northwestern University, Evanston, IL 60208, USA}	
	\affiliation{Department of Mechanical Engineering, Northwestern University, Evanston, IL 60208, USA} 
	\affiliation{Northwestern Institute on Complex Systems (NICO), Northwestern University, Evanston, IL 60208, USA}
	
	\author{Richard M. Lueptow}
	\affiliation{Department of Mechanical Engineering, Northwestern University, Evanston, IL 60208, USA}
	\affiliation{Department of Chemical and Biological Engineering, Northwestern University, Evanston, IL 60208, USA}
	\affiliation{Northwestern Institute on Complex Systems (NICO), Northwestern University, Evanston, IL 60208, USA}

	\begin{abstract}
		%The segregation patterns generated by granular motion of particles of different sizes in a fully 3D flow --- a spherical tumbler rotated alternately about two perpendicular axes --- are studied in terms of experiments, both direct observation at the tumbler wall and x-ray imaging, and a continuum model of the flow from which Poincar\'{e} sections can be found. Experiments span a range of volume fractions of each particle species and size ratios. The mechanism of pattern formation is examined by relating particle size segregation to the chaotic flow and non-mixing regions evident from the continuum model based on flow kinematics only, which is connected with a mathematical model based on piecewise isometries, under the framework of cutting and shuffling, a paradigm of mixing discrete materials. Alignment of non-mixing periodic islands of the underlying flow field with segregation due to particle size difference in multiple dimensions enables the accumulation of a single particle species in the non-mixing islands against collisional diffusion and chaotic transport. Flow behaviors surrounding the non-mixing islands embodied by  dynamical systems approach also provide evidence that some segregation patterns are more robust than others.
		
		Segregation patterns of size-bidisperse particle mixtures in a fully-three-dimensional flow produced by alternately rotating a spherical tumbler about two perpendicular axes are studied over a range of particle sizes and volume ratios using both experiments and a continuum model.  Pattern formation results from the interaction of size segregation with chaotic regions and non-mixing islands of the flow.  Specifically, large particles in the flowing  surface layer are preferentially deposited in non-mixing islands despite the effects of collisional diffusion and chaotic transport. The protocol-dependent structure of the unstable manifolds of the flow surrounding the non-mixing islands provides further insight into why certain segregation patterns are more robust than others. 
	\end{abstract}

	\maketitle
	
	\section{Introduction}
 	
 	Segregation of flowing granular materials, differing in properties such as density or size, has been studied both experimentally and theoretically in a number of canonical geometries including chutes \cite{Savage1988,Dolgunin1995}, quasi-two-dimensional (2D) bounded heaps \cite{Fan2017}, and annular shear cells \cite{Golick2009}, where the underlying flow field is relatively simple and develops easily predicted segregation patterns. However, the situation can be more complex when the chaotic dynamics of the underlying flow field interacts with segregation, as can occur in quasi-2D tumblers \cite{Hill1999,Fiedor2005,Khakhar1999,Meier2006,Meier2007}. Although chaotic flows have been well studied in fluids \cite{Haller2000,Ottino1989}, similar studies with granular systems are few, particularly for three-dimensional (3D) systems. 
 	
 	As an example of the interaction between segregation and chaotic dynamics, consider the case of a quasi-2D tumbler with square cross-section rotated at a constant speed \cite{Meier2006}, where non-trivial segregation pattern forms due to the competing influences of segregation and the underlying flow. Here, the tumbler is initially half filled with a uniform mixture of small (diameter $d$ = 0.3 mm) black glass particles  and large ($d$ = 1.2 mm) clear glass particles.
	\begin{figure}[h]
		\includegraphics[width=0.8 \columnwidth]{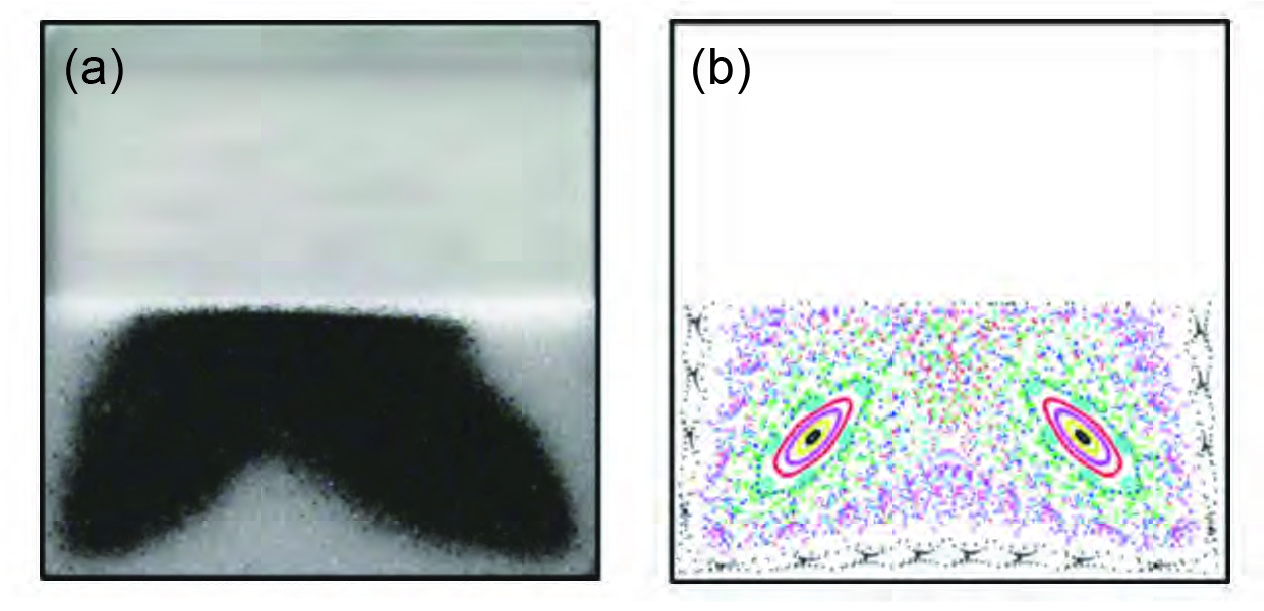}
		\caption{\label{Fig:1} (a) Segregation experiment in a half-full square tumbler with $40\%$ small (0.3 mm) black particles and $60\%$ large (1.2 mm) clear particles by weight. Steady-state pattern after ten clockwise revolutions of the tumbler at 1.44 rpm. (b) Poincar\'{e} section of a half-full square tumbler derived from model of flow kinematics. Reprinted with permission from Meier et al.~\cite{Meier2006} \copyright 2006 American Physical Society. }
	\end{figure}
	The tumbler is rotated at constant angular speed so that particles continuously flow down the free surface (rolling/cascading regime \cite{Mellmann2001,Henein1983}). After several revolutions, shown in Fig.~\ref{Fig:1}(a), the small black particles accumulate in the two lobes,  which extend from the core toward the two corners, while large clear particles occupy the periphery of the tumbler. The lobed pattern comes from the time-periodic nature of the flow due to the tumbler geometry. That is, the surface flowing layer varies in length periodically from the position shown in Fig.~\ref{Fig:1}(a) to a position where it spans the diagonal of the tumbler. The corresponding Poincar\'{e} section \cite{Strogatz2001}, a stroboscopic mapping of points advected by a simple kinematic model of the granular tumbler flow, is shown in Fig.~\ref{Fig:1}(b). The key point here is that the Poincar\'{e} section is based on a continuum model derived purely from the velocity field, without any information concerning the particles used in the experiments or their tendency to segregate. Nevertheless, the correspondence between the two non-mixing elliptic islands along the diagonals (evident as colored ellipses and associated lobes) in the Poincar\'{e} section and the lobes of small black particles in the experiment is clear. The segregation pattern is a manifestation of the interplay between segregation of particles in the flowing layer and the dynamics of the time-periodic flow in the tumbler. Segregation drives small particles to percolate to the bottom of the flowing layer where they are influenced by the advection of the underlying flow field to accumulate in the non-mixing elliptic islands.

	While the interaction between particle segregation in the quasi-2D square tumbler in Fig.~\ref{Fig:1}(a) and the 2D chaotic dynamics of the associated Poincar\'{e} section based only on the kinematics of the flow is quite evident, it is not obvious if the same interaction will occur in a fully 3D system. In this paper, we examine whether or not granular segregation and chaotic dynamics can interact in a similar way in a 3D system so as to generate segregation patterns related to  non-mixing regions. 
	
	\begin{figure}
		\includegraphics[width= 0.6\columnwidth]{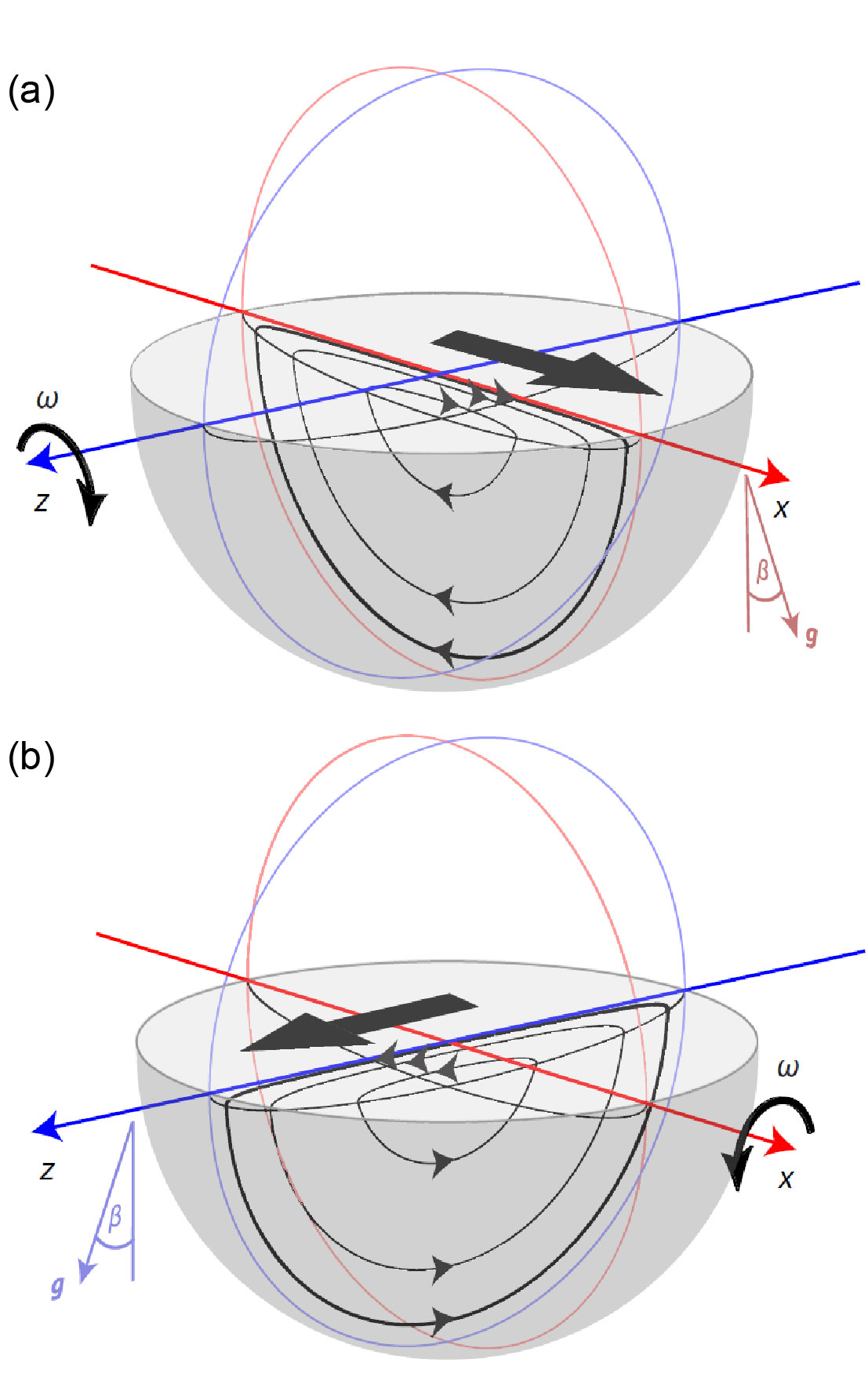}
		\caption{\label{Fig:2}Biaxial spherical tumbler flow consists of two alternating single axis rotations about (a) the $z$-axis and (b) the $x$-axis, at a rotation speed $\omega$. Material passing through the flowing layer (with a flat free surface at angle $\beta$ to horizontal) is subsequently deposited downstream and then moves in solid-body rotation with the tumbler until it again enters the flowing layer. Reprinted with permission from Zaman et al. \cite{Zaman2018} \copyright 2018 Nature.}
	\end{figure}

\begin{figure}[h]
		\includegraphics[width=0.8\columnwidth]{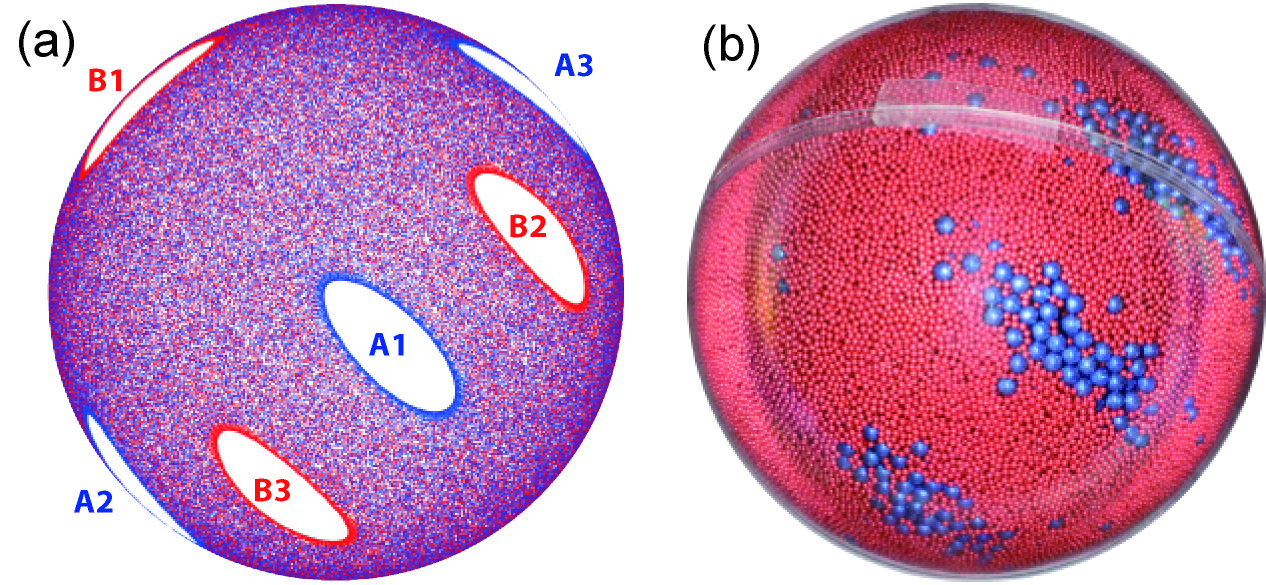}
		\caption{\label{Fig:3}(a) Bottom view of the Poincar\'{e} section of a half-full spherical tumbler for $r$ = 0.95 with protocol (57$^{\circ}, 57^{\circ}, 90^{\circ}$). (b) Bottom view of a segregation experiment in a half-full spherical tumbler with 15\% large ($d$ = 4 mm) blue particles and 85\% small ($d$ = 1.5 mm) red particles by volume for the same protocol as in (a). The pattern forms after 30 iterations of protocol. Red and blue colors in experiments and in the continuum model do not correspond with one another.}
	\end{figure}

	To answer this question, we consider segregation patterns in a 3D spherical tumbler that is half-filled with a mixture of small and large millimeter-sized spherical particles. As shown schematically in Fig.~\ref{Fig:2}, the tumbler is rotated by angle $\theta_z$ about the $z$-axis and then by angle $\theta_x$ about the $x$-axis, where the $z$-axis and $x$-axis both lie in the horizontal plane. In the figure and in most previous studies, the axes are orthogonal, but in general, the angle $\gamma$ between them can have any value \cite{Juarez2012,Lynn2018}. This biaxial rotation protocol is specified by the triple ($\theta_z,\theta_x,\gamma$) and typically is repeated many times. The pattern formation presented in this work depends on the specific protocol angles, mixture compositions, and particle sizes. In fact, there are other factors that can influence the result by contributing in similar ways to the mechanism, which for simplicity are out of the scope of this study. Christov et al. \cite{Christov2014} identified chaotic mixing regions coexisting with regular non-mixing islands based on a continuum model calculation of the underlying flow for certain protocols. An example of the coexisting chaotic region and non-mixing islands in the Poincar\'{e} section is shown in the bottom view of the system in Fig.~\ref{Fig:3}(a), where the white elliptical regions (labeled A1-A3 and B1-B3) are non-mixing islands surrounded by the chaotic region covered in blue and red tracer points. This Poincar\'{e} section is calculated from the continuum model outlined in Christov et al. \cite{Christov2014} by tracing points initially seeded on the interface between the bulk and the flowing layer [see Appendices \ref{CM} and \ref{PS}]. The red and blue colors in the continuum model indicate tracer points for the $z$-axis and $x$-axis actions, respectively. 
	
	More recently, Zaman et al. \cite{Zaman2018} experimentally demonstrated for monodisperse particles that elliptic non-mixing islands in the flow  serve as barriers to mixing by prohibiting material exchange across their boundaries. 
	Specifically, they showed that a single tracer particle can stay within the non-mixing regions near the tumbler wall, periodically appearing in each of the period-3 non-mixing regions (either A1-A2-A3 or B1-B2-B3), over hundreds of iterations of the protocol in a spherical tumbler rotated about two orthogonal axes. Occasionally, the tracer particle wanders into the chaotic region due to collisional diffusion, but it eventually returns to the non-mixing regions where it can again remain hundreds more iterations. These non-mixing regions also can be predicted by the more abstract mathematical theory of piecewise isometries (PWI) \cite{Goetz1998,Goetz2000}, where discontinuities can generate complex dynamical behaviors as seen in various applications \cite{Sturman2012,Deane2006}. The PWI map, which applies to the limiting case of an infinitely thin flowing layer at the free surface \cite{Smith2017,Park2016,Juarez2010}, captures the skeleton of the underlying flow generated by the fundamental framework of cutting-and-shuffling, a mechanism for mixing discrete materials \cite{Christov2011,Juarez2010,Juarez2012,Lynn2018,Zaman2018,Yu2016}. 
	
	In this paper, we replace the monodisperse particles and single large tracer particle used in  our previous work \cite{Zaman2018} with a mixture of small and large particles to explore if the segregation pattern in a fully 3D system matches the prediction of the Poincar\'{e} section derived from the flow kinematics alone. This would be  analogous to the quasi-2D segregation pattern in Fig.~\ref{Fig:1}(a) matching the 2D Poincar\'{e} section in Fig.~\ref{Fig:1}(b). Apart from the fundamental question of whether 3D chaotic dynamics interact with granular segregation, this research has implications for practical devices for mixing granular materials in which non-mixing (segregation) regions are detrimental to the mixing process, a critical issue in mixing powders in the pharmaceutical and chemical industries. 
	
	The answer to the question of whether or not granular segregation and chaotic dynamics can interact in a 3D system to generate segregation pattern is immediately evident from the bottom view of an experiment with a mixture of initially mixed large blue particles ($d$ = 4 mm) and small red particles ($d$ = 1.5 mm) for protocol (57$^{\circ}, 57^{\circ}, 90^{\circ}$), shown in Fig.~\ref{Fig:3}(b) after 30 iterations. Large blue particles accumulate in the period-3 non-mixing regions at the bottom center of the tumbler, corresponding to A1, B2, and B3 islands in the Poincar\'{e} section in Fig.~\ref{Fig:3}(a). The red and blue colors in the continuum model indicate tracer points for the $z$-axis and $x$-axis actions, respectively. In the next section we will show that large blue particles also accumulate in the non-mixing regions labeled as A2, A3, and B1 on the periphery of the tumbler, which is not visible in Fig.~\ref{Fig:3}(b) due to curvature of the tumbler surface.  
	
	Thus, it appears that the same mechanism that leads to pattern formation in the quasi-2D system of Fig.~\ref{Fig:1} also occurs in a fully 3D system. That is, segregation due to particle size difference effectively drives one particle species into the non-mixing features derived from dynamical system models. In the remainder of this paper we explore pattern formation in the 3D spherical tumbler, its relation to chaotic dynamics, and the details of the mechanisms that are involved.

 	\begin{figure*}[ht]
		\includegraphics[width=0.8\textwidth]{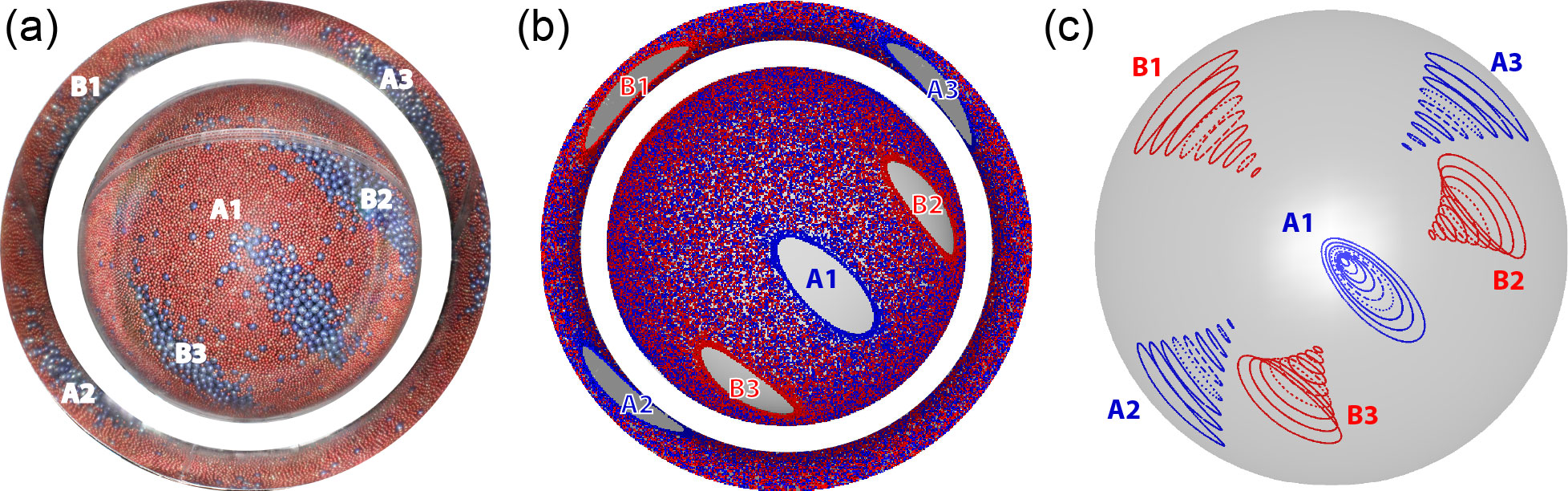}
		\caption{\label{Fig:4}(a) Segregation experiment in a half-full spherical tumbler rotated at 2.6 rpm with $10\%$ large blue particles (3 mm) and $90\%$ small red particles (1 mm) by volume showing six distinct large particle islands for protocol (57$^{\circ}, 57^{\circ}, 90^{\circ}$). The outer ring is the tumbler reflection in the surrounding cylinder, which shows the periphery of tumbler. (b) POV-Ray \cite{povray} generated view of Poincar\'{e} section including the  reflective cylinder. Flowing layer edges are tracked for 500 iterations near the tumbler wall at nondimensional radius $r$ = 0.95. (c) Bottom view of 3D non-mixing islands from continuum model for $0.55 < r < 0.95$ in increments of 0.05 consist of a pair of period-3 islands for a total of six conical-shaped non-mixing structures (A1-A2-A3 and B1-B2-B3).  }
	\end{figure*}

	\section{Segregation Pattern Visualization}

	Experiments are conducted using the same apparatus and methodology described previously \cite{Zaman2018}. An acrylic spherical tumbler of diameter $D = 2R_0 = 14$\,cm half-filled with millimeter-sized spherical glass particles is placed on an apparatus consisting of a set of three rollers mounted on a turntable. The spherical tumbler rests on the rollers, one of which is driven by a motor to rotate the tumbler about a single horizontal axis at 2.6 rpm. To rotate the tumbler about an orthogonal horizontal axis, a mechanism lifts the tumbler off the rollers, the turntable is rotated, and the tumbler is set back down on the re-oriented rollers. In this way, the tumbler can be repeatedly rotated about two horizontal axes in an alternating fashion to perform a biaxial rotation protocol. In previous work \cite{Zaman2018}, only a single large tracer particle was tracked. Here, we use a large number of larger diameter particles so that segregation patterns can form.

	To qualitatively analyze the segregation pattern, we photograph the tumbler from below. After a desired number of biaxial rotations, the tumbler is removed from the  apparatus,  placed in a hole in a metal plate having a diameter slightly smaller than that of the tumbler, and photographed from below. Due to the curvature of the spherical tumbler, only the lower portion of the hemisphere is clearly captured in the photo [center circular image in Fig.~\ref{Fig:4}(a)]. To view the periphery of the tumbler, a polished circular aluminum cylinder (15.2 cm diameter by 7.6 cm long) is attached to the bottom of the plate concentric with the tumbler. In this way, the reflection of the periphery of the tumbler is also captured in the photo [ring in Fig.~\ref{Fig:4}(a)]. The tumbler is illuminated by a point LED light source positioned just to the side of the camera lens. Two photos are taken with the light source on the left and right sides of the camera lens while maintaining the same camera location and settings. The final image is obtained by combining left and right halves of the two photos that do not have glare or shadows.  
	The light-colored arc above the label  B2  in Fig.~\ref{Fig:4}(a) is the seam between the two halves of the clear spherical tumbler.
	
	Figure~\ref{Fig:4}(a) shows pattern formation for the protocol ($57^{\circ}, 57^{\circ}, 90^{\circ}$). Here, large blue particles accumulate in non-mixing regions surrounded by small red particles.  There are six clusters of large blue particles, three in the center circle (A1, B2, and B3) and three reflected in the ring (A2, A3, and B1). These clusters  align closely with the six non-mixing islands evident in the Poincar\'{e} section in Fig.~\ref{Fig:4}(b), which is constructed according to the process outlined in Appendices \ref{CM} and \ref{PS}. The tumbler and the reflected ring in the Poincar\'{e} section are constructed using POV-Ray \cite{povray} in the same manner the tumbler is photographed in Fig.~\ref{Fig:4}(a). 
	
	The remarkable similarity between the segregation pattern of clusters of large particles in Fig.~\ref{Fig:4}(a) and the elliptic (non-mixing) regions in the Poincar\'{e} section in Fig.~\ref{Fig:4}(b) is a key result of this paper. That is, the Poincar\'{e} section, which is based only on a simple kinematic model of the velocity in the flowing surface layer in the tumbler and has no particle segregation model at all, accurately predicts the regions in which segregating particles (large particles in this case) accumulate. Thus, the chaotic dynamics of the system, as represented by the Poincar\'{e} section, predicts the segregation pattern. 
	
	Of course, unlike the quasi-2D systems shown in Fig.~\ref{Fig:1}, Poincar\'{e} sections in a fully 3D systems are also three dimensional. Figures~\ref{Fig:3} and \ref{Fig:4} only show the segregation pattern visible at the clear wall of the spherical tumbler and the corresponding Poincar\'{e} section at a dimensionless radius of $r$ = 0.95 (just adjacent to the wall of the tumbler). These 2D non-mixing islands on different invariant surfaces form a 3D structure around a line of periodic points \cite{Pouransari2010,Mullowney2005,Moharana2013,Christov2014,Meier2007}. To visualize the 3D structure in the spherical tumbler flow, trajectories of the points on the boundaries of non-mixing islands for the ($57^{\circ}$,~57$^{\circ}$,~90$^{\circ}$) protocol in the radius range of $0.55 < r < 0. 95$ are isolated and assembled in Fig.~\ref{Fig:4}(c) to form six conical Kolmogorov-Arnold-Moser (KAM) tubes \cite{Christov2014,Pouransari2010}, analogous to 2D KAM islands. The shape of the 3D KAM structure is determined by the protocol. The 3D KAM structures shown here for the ($57^{\circ}$,~57$^{\circ}$,~90$^{\circ}$) protocol are conical with their apex pointing toward the center of the hemisphere. The bases of the conical KAM tubes are at the wall of the tumbler and correspond to the islands A1-A3 and B1-B3 in which the large particles accumulate in Fig.~\ref{Fig:4}(a). Viewing the hemisphere from the bottom as in Fig.~\ref{Fig:4}(c), the A2, A3, and B1 KAM tubes are viewed from the side making their conical shape obvious, while the A1 KAM tube is viewed from its base [leading to the nested ellipses in Fig.~\ref{Fig:4}(c)]. KAM tubes B2 and B3 are between these extremes of orientation. The largest ellipse in each of the six KAM tubes, which is the portion of the KAM tubes closest to the wall of the tumbler, corresponds to the elliptical non-mixing regions in Figs.~\ref{Fig:4}(a) and \ref{Fig:4}(b).
	
		\begin{table*}	
		\begin{tabular}{ c | c | c | c | c | c | c  }
			\hline
			\hline
			color & \multicolumn{4}{|c} {red} & \multicolumn{2}{|c} { blue} \\ \hline
			nominal size (mm) & 1 & 1.5 & 2  &  3&2 & 4\\ \hline 
			actual size (mm) &1.12 $\pm$ 0.09&1.50 $\pm$ 0.13& 1.90 $\pm$ 0.09& 3.15 $\pm$ 0.14 & 2.03 $\pm$ 0.12&3.99 $\pm$ 0.04\\ \hline
			\hline		
		\end{tabular}
		\caption{\label{Tab:2} Nominal particle sizes and corresponding actual particle sizes.  }
	\end{table*}

\begin{figure*}[htb]
		\includegraphics[width=0.7\textwidth]{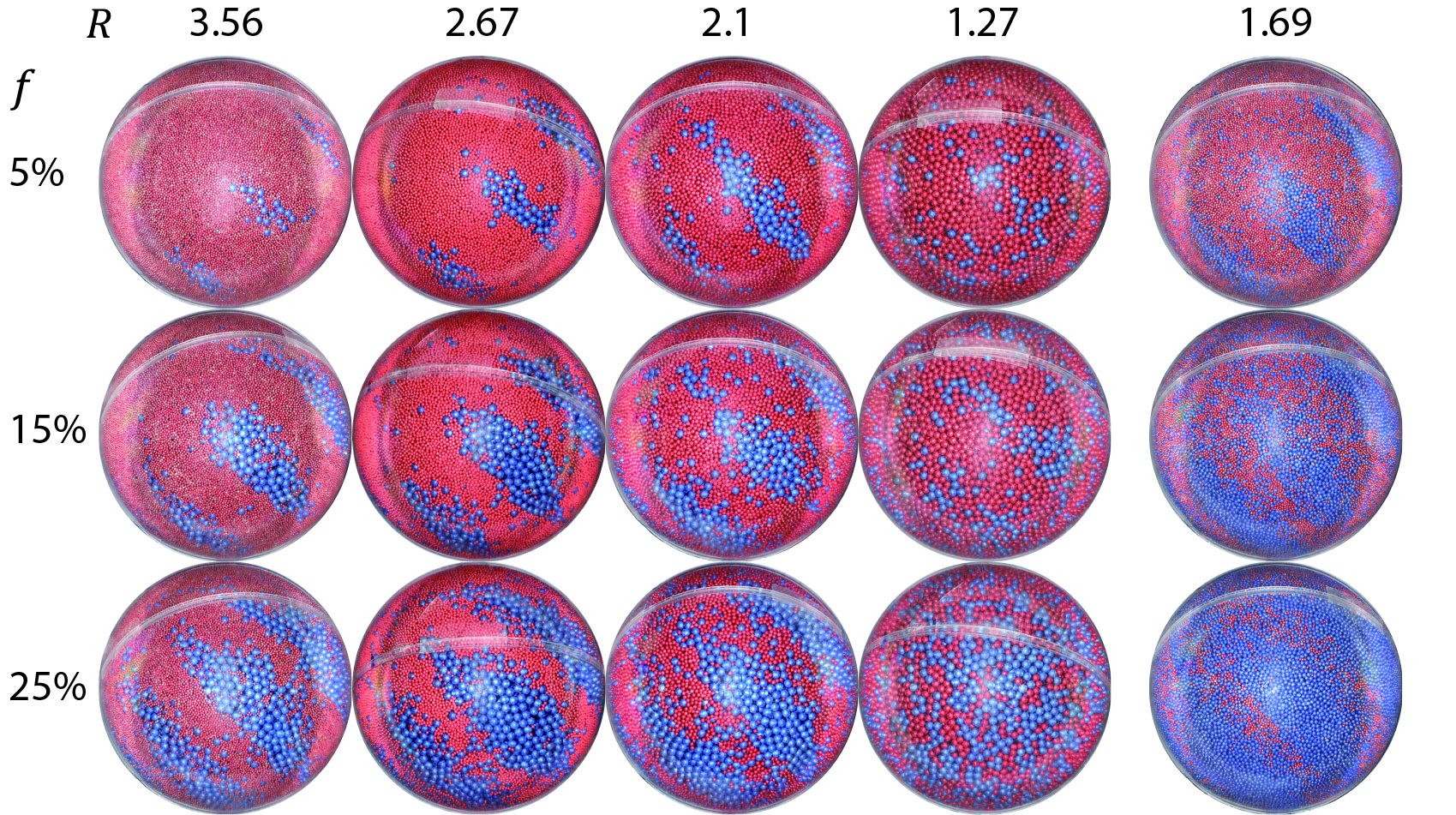}
		\caption{\label{Fig:5}Segregation experiments in a half-full spherical tumbler with protocol  ($57^{\circ}, 57^{\circ}, 90^{\circ}$) rotated at 2.6 rpm for a range of mixtures: $5\%$, $15\%$, and $25\%$ volume fraction $f$ of large blue particles, for size ratios $R$  = 3.56, 2.67, 2.1, 1.27, 1.69. All columns use large particles with 4 mm diameter except for the rightmost column where the large particle diameter is 2 mm.}
	\end{figure*}

	In Fig.~\ref{Fig:4}(a), the mixture consists of $10 \%$  large (3 mm) blue particles and $90 \%$ small (1 mm) red particles. The large blue particles accumulate into regions that correspond to non-mixing islands predicted by the continuum model. To fully characterize the pattern formation, experiments are carried out across a range of particle size ratios $R$ and large particle volume fractions $f$. Figure \ref{Fig:5} shows experimental results for mixtures of equal-density spherical glass particles with actual particle size ratios 3.55, 2.67, 2.1, 1.27, and  1.69. Nominal  and actual particle diameters of the mixtures used in the experiments are listed in Table~\ref{Tab:2}. The volume fraction  of large particles is 5\%, 15\%, and 25\%. In the first four columns of Fig.~\ref{Fig:5}, the large particle size is kept constant at 4 mm to allow a direct visual comparison of the coverage of large particles on the tumbler wall. The rightmost column with $R$  = 1.69 and a large particle diameter of $d$ = 2 mm provides a comparison to size ratio $R$  = 2.1 images with  large particle diameter of $d$ = 4 mm to assess the effect of the tumbler size relative to particle sizes.

	Consider the first row of experiments with  $f$ = 5\%. The three clusters of blue particles match with non-mixing islands  A1, B2, and B3  predicted by the Poincar\'{e} section calculated from the continuum model in Fig.~\ref{Fig:4}(b). The boundaries of the regions of blue particles become more difficult to discern as the particle size ratio decreases, particularly for $R$ = 1.27. This result is expected since a smaller size ratio leads to weaker segregation between large and small particles. Consequently, particles have a greater tendency to remain in a mixed state, resulting in less distinctly segregated regions with a smaller size ratio. It is also evident that the islands of blue particles are smaller for larger size ratios. This is likely a result of the relatively small fraction of large particles and the large particle size ratio, which allows small particles to populate regions at the wall below the large particles.
	
	As the volume fraction of large particles increases (rows two and three in Fig.~\ref{Fig:5}), the same segregation pattern persists. For all size ratios, the clusters of large particles occupy larger area as $f$ increases from 5\% to 25\%, growing in both length and width, because more large particles are available to accumulate in the non-mixing regions. In general, accumulation of large particles into the non-mixing regions is more distinct with a larger size ratio, while collisional diffusion and segregation make the patterns less evident with decreasing size ratio so that for a size ratio of $R$ = 1.27 and $f$ = 25\% the segregation pattern is no longer evident. 
	
	For the four size ratios on the left in Fig.~\ref{Fig:5}, the large particle size remains at 4 mm while the small particle size varies. Consider now the size ratio $R$ = 1.69 in Fig.~\ref{Fig:5}, where the large particle size is now 2 mm instead of 4 mm. The accumulation of large blue particles into three non-mixing regions is still observed across the different large particle volume fractions. At volume fraction 25\%, compared with a similar size ratio $R$ = 2.1, the coverage of large blue particles is much higher with fewer small red particles visible. Yet, it is still obvious that large blue particles accumulate more intensely in the non-mixing regions.	
	
	The series of experiments in Fig.~\ref{Fig:5} demonstrates that the segregation pattern for this particular protocol ($57^{\circ}, 57 ^{\circ}, 90 ^{\circ}$) is quite robust and manifests itself across a wide range of particle size ratios and large particle volume fractions, despite the interplay between collisional diffusion and segregation due to particle size difference. Collisional diffusion and chaotic advection tend to disperse the particles throughout the domain. At the same time, the segregation between large and small particles traps many large particles in the non-mixing regions resulting in the pattern formation.

	\section{Non-mixing structures in 3D }
	Visualization of the segregation patterns formed at the tumbler wall in the previous section is an intuitive way to qualitatively analyze the pattern formation, but only near the tumbler wall. As shown in Fig.~\ref{Fig:4}(c), the non-mixing structures are conical-shaped in three dimensions. Hence, x-ray imaging is employed to examine the 3D nature of particle segregation. 
	\begin{figure}[h]
		\includegraphics[width=0.5\columnwidth]{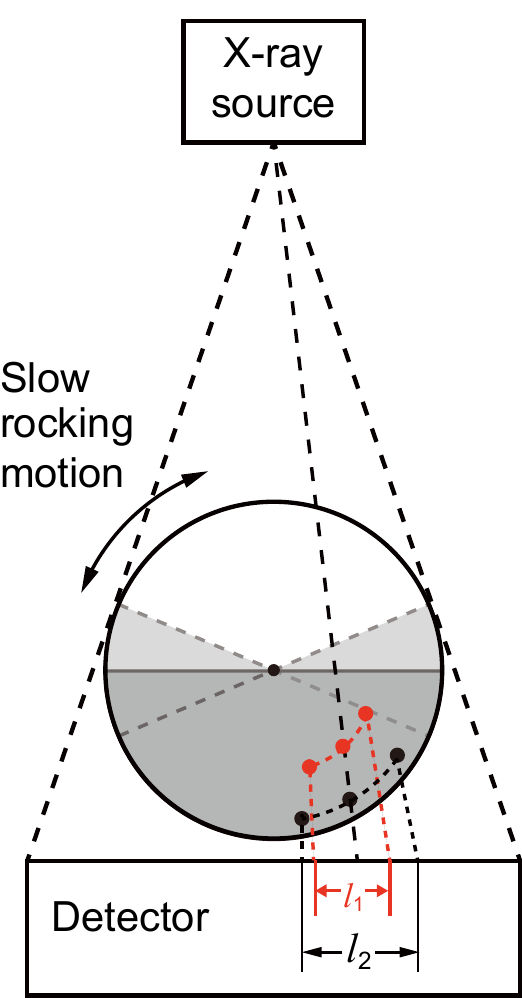}
		\caption{\label{Fig:6} Schematic of depth determination. Cone-shaped x-ray beam passes through the tumbler and x-ray images are captured by a detector below the tumbler. Rotating the tumbler between the two extremes defined by the repose angle results in trajectories $l_{1}$ and $l_{2}$ at the detector that depend on the radial location of the particle in the tumbler. }
	\end{figure}
	
	The x-ray imaging equipment used in this study is the same as used previously \cite{Zaman2018}.
	X-ray images of the tumbler are captured by a detector located below the tumbler apparatus as shown in the schematic in Fig.~\ref{Fig:6}. The cone-shaped x-ray beam makes it possible to obtain the 3D positions of multiple tracer particles as follows. When the tumbler is slowly rotated to the granular material's angle of repose and back, all particles move in solid body rotation. A sequence of images recording positions of x-ray opaque tracer particles is captured, from which the 3D trajectories of particles can be tracked. These trajectories project onto the detector image at different trajectory lengths depending on the particle depth, as shown in the schematic side view in Fig.~\ref{Fig:6}. For example, the black particle that is closer to the tumbler wall has a longer projected trajectory ($l_{2}$) than that of the red particle ($l_{1}$), which is located further radially inward. However, the tracer particle positions are projection-distorted because the x-ray point source rays that land on the flat detector traverse different distances depending on the depth of the particle and how far it lies from the source-detector central axis. The 3D position of a tracer particle can be calculated iteratively from the sequence of the 2D images during the slow rotation with particles moving in solid body rotation (a rocking motion). Based on the tracer particle coordinates in the images, an approximate radial location of the particle in the tumbler can be estimated from the trajectory during the rocking. With the estimated radius, coordinates of the trajectory are adjusted accordingly, and a new value of the radius can be calculated. This process is repeated until the difference between successive iterations for the radius location in the tumbler is less than $1\%$. With multiple tracer particles, a particle tracking velocimetry (PTV) algorithm in MATLAB is used to detect and trace particle trajectories \cite{Crocker1996}. In this way, the 3D position of many x-ray opaque tracer particles can be obtained before and after each rotation of the protocol.
	
	The captured image is first calibrated to remove warping effects in image detection. The actual tumbler is 14 cm in diameter and appears as a circular image with a diameter of 612 pixels. The tracer particle used in tracking experiments is about 3 mm in diameter, which is about 13 pixels. Images are taken one frame per angle of rotation, which results in a 23 image sequence for a single side rotation to static angle of repose ($\beta = 23^{\circ}$). For experiments of rotating $\pm\beta$, only trajectories longer than 30 frames are recorded to ensure more accurate tracking. This minimum tracking criterion will also reduce error for particles overlapping in the detection images.
	
	The apparatus and analysis methods have some limitations. Tracking the trajectories of tracer particles requires accurate detection of the particles in the x-ray detector image. When a large number of tracer particles is used, particles often overlap in the x-ray image. Particles at various depths may be projected to the same neighborhood on the detector image, particularly as particles accumulate into a non-mixing region and overlap. At the same time, when the tumbler is rotated to the angle of repose, the relative particle locations change, so that particles may cross over each other or overlap in the images. 
	Thus, it is challenging to track a large number of particles. To avoid inaccurate detection of particles, experiments are limited to fewer than 200 x-ray opaque tracer particles. With 4 mm particles, this comprises less than $2\%$ of the total particle volume. 
	Thus, for experiments with volume fractions of large particles higher than $2\%$, both large x-ray opaque tracer particles and large glass beads of similar density are used to provide the appropriate volume of large particles. Nevertheless, the tracer particles still represent the ensemble behavior of all large particles.

	The continuum model for the ($57^{\circ},57^{\circ}, 90^{\circ}$) protocol predicts conical non-mixing structures, each with its base on the wall of the spherical tumbler and its apex pointing toward the center of the tumbler [Fig.~\ref{Fig:4}(c)]. The apices of the conical non-mixing structures reach the size of a typical large particle at a dimensionless radius of about $r$ = 0.55 . If the particle is larger than the KAM tube itself, a single particle in a non-mixing region cannot be distinguished from particles dispersed randomly in the chaotic region. Therefore, non-mixing structures in the experiment should only occur for locations in the tumbler for $r >$ 0.55. We use this result to explore how particles segregate and patterns form in the tumbler. Two experiments tracking x-ray opaque tracer particles are performed, first targeting particles that start near the center of the bed of particles but should segregate to near the tumbler wall and accumulate in non-mixing regions, and second, targeting particles that start near the tumbler wall but should segregate toward  the center of the tumbler, to the portion of the bed where the conical non-mixing regions do not extend. 

		\begin{figure}[h]
		\includegraphics[width= 0.8\columnwidth]{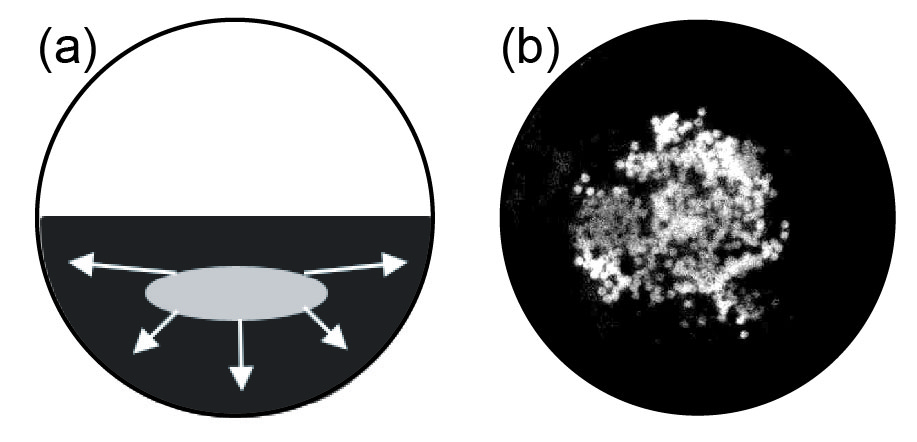}
		\caption{\label{Fig:7}Initial condition for tracking particles segregating toward the tumbler wall. Large x-ray opaque 3 mm silver particles (gray blob) are seeded in the core of the hemispherical bed of 1.8 mm glass particles (black background) as shown in (a) a side view schematic with arrows indicating that these particles are expected to segregate toward the tumbler wall  and (b) a bottom-view x-ray image.}
	\end{figure}
	
	\begin{figure}[h]
		\includegraphics[width= 0.9\columnwidth]{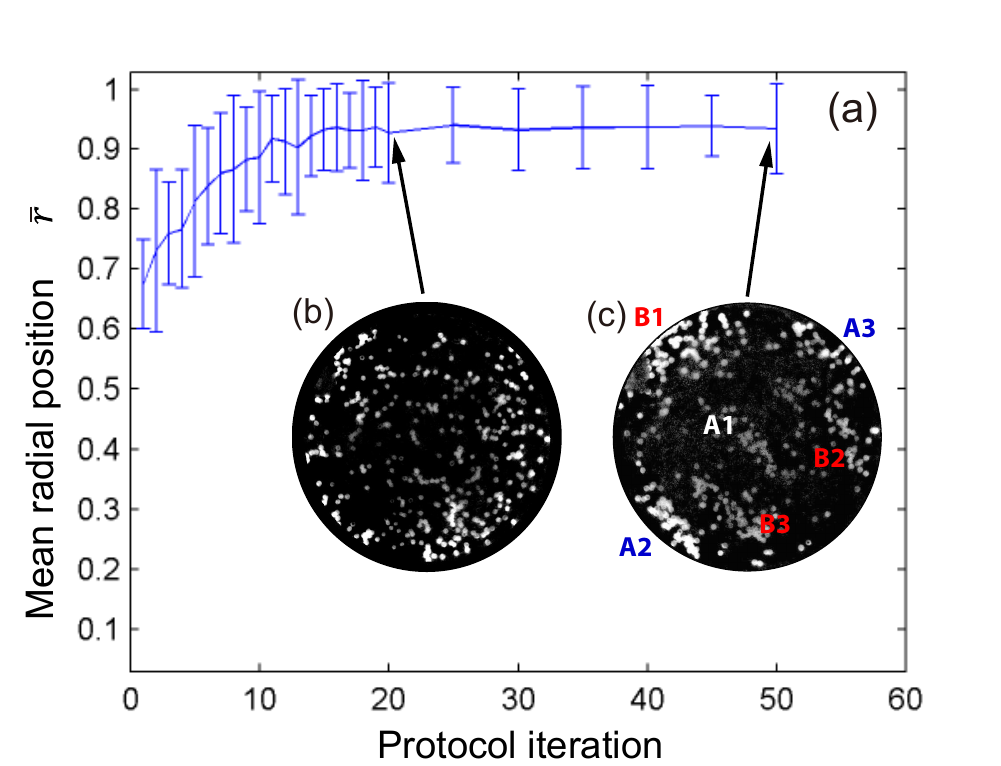}
		\caption{\label{Fig:8} (a) Mean normalized radius of large x-ray opaque tracer particles is plotted as a function of protocol iterations for protocol ($57^{\circ}, 57^{\circ}, 90^{\circ}$). X-ray image of tracer particles at (b) 20 iterations and (c) 50 iterations. Particles have sergegated to near the tumbler wall after 20 iterations, but only at 50 iterations are the tracer particles segregated into non-mixing islands labeled by A1-A2-A3 and B1-B2-B3. Error bars are one standard deviation.}
	\end{figure}

	In the first experiment, 400 $d = 3.01\pm0.04$\, mm, $\rho  = 2.5\pm0.05$\, g cm$^{-3}$ hollow silver jewelry beads (Beadcorp) with 0.9 mm holes are used as x-ray tracer particles with $d = 1.84 \pm 0.07$ mm $\rho = 2.45 \pm 0.3$\, g cm$^{-3}$  glass particles in the bulk. The tracer particles are seeded in the core of the tumbler [Fig.~\ref{Fig:7}(a)], appearing as a circular cluster in the bottom-view x-ray image [Fig.~\ref{Fig:7}(b)].  Thus, the initial blob of large x-ray opaque tracer particles starts at a radial location near the apices of the conical non-mixing structures. Of course, once the rotation protocol starts, the large x-ray opaque particles should migrate to near the wall of the tumbler due to segregation in the flowing layer, recalling that larger particles preferentially segregate to the surface of the flowing layer and thereby to the periphery of the bed of particles \cite{Nityanand1986}. The question is: Do the x-ray opaque particles accumulate in non-mixing regions before or after they get near the wall of the tumbler? 
	
	To answer this question, in this experiment, after each iteration, the 3D positions of all tracer particles are obtained. The mean radius $\bar{r}$ of tracer particles in the tumbler is calculated to quantify the extent of the segregation. In Fig.~\ref{Fig:8}(a), the normalized mean radius of tracer particles $\bar{r}/R_0$ is plotted against the number of iterations. The mean radius starts around 0.7, increases rapidly in the first 20 iterations, and then stays fairly constant at about 0.9. Thus, the tracer particles segregated to near the wall of the tumbler, as expected. At 20 iterations [Fig.~\ref{Fig:8}(b)], tracer particles are dispersed randomly throughout the image but are near the tumbler wall based on Fig.~\ref{Fig:8}(a). After 50 iterations [Fig.~\ref{Fig:8}(c)], tracer particles form clusters in the non-mixing regions as predicted by the continuum model, labeled as A1-A2-A3 and B1-B2-B3.
	Note that the the intensity of this image is adjusted using adaptive histogram equalization in MATLAB (\texttt{adapthisteq}) \cite{Pizer1987,Zuiderveld1994}, where the contrast is enhanced in  smaller tiles across the entire image. 
	The clusters form slowly, becoming more evident toward the end of the experiment. This result demonstrates that large particles accumulate in non-mixing regions consistent with the continuum model but only after they have first segregated to near the wall of the tumbler where the non-mixing regions have significant volume due to their conical shape.

	\begin{figure}[h]
		\includegraphics[width=0.8\columnwidth]{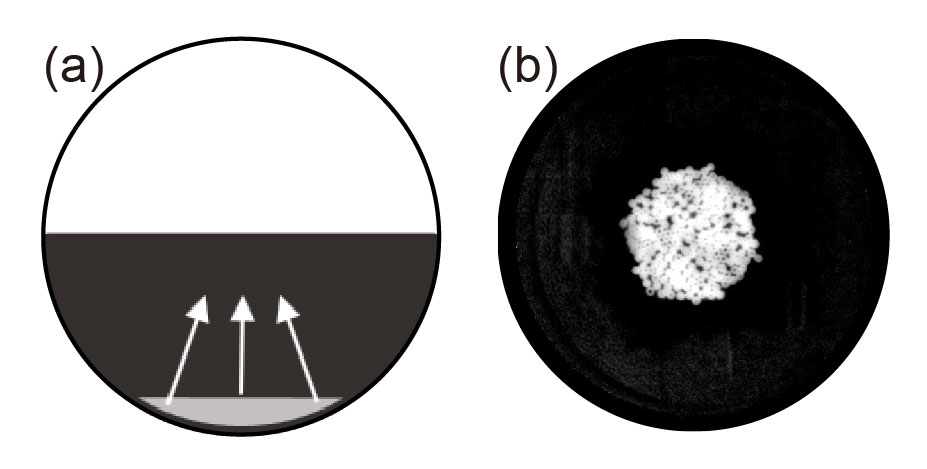}
		\caption{\label{Fig:9}Large silver particles (3 mm) are seeded in the bottom of the hemispherical bed of 2 mm acrylic particles as shown in (a) the side view schematic with arrows indicating that these particles are expected to segregate toward the center of the particle bed and (b) a bottom-view x-ray image.}
	\end{figure}
	
	\begin{figure}[h]
		\includegraphics[width=0.9\columnwidth]{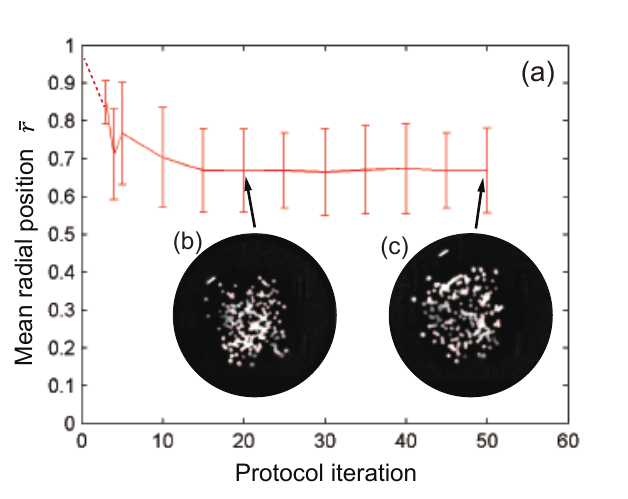}
		\caption{\label{Fig:10} (a) Mean normalized radius of x-ray opaque tracer particles is plotted as a function of number of protocol iterations ($57^{\circ},57^{\circ}, 90^{\circ}$). X-ray image of tracer particles at (b) 20 iterations and at (c) 50 iterations. Error bars are +/- one standard deviation.}
	\end{figure}

	In the second experiment we use a mixture of 400  $d$ = $3.01\pm0.04$ mm hollow silver jewelry beads ($\rho  = 2.5\pm 0.05$\, g cm$^{-3} $) as tracers with $d$ = $1.92 \pm 0.04$ mm acrylic particles having a density $\rho = 1.24 \pm 0.08$\, g cm$^{-3}$ in the bulk. This combination of size and density differences causes the silver particles to sink to the bottom of the flowing layer during tumbler rotation while the glass particles rise to the surface of the flowing layer so they segregate toward the tumbler wall. The silver particles are seeded at the bottom of the tumbler as shown in the side view and bottom view in Figs.~\ref{Fig:9}(a) and \ref{Fig:9}(b), respectively. Since silver particles sink to the bottom of the flowing layer, they are expected to segregate toward the core of the hemispherical bed of acrylic particles as the tumbler is rotated. The normalized mean radius of silver particles is plotted as a function of the number of protocol iterations in Fig.~\ref{Fig:10}(a). Due to difficulties detecting the silver particles in the first few iterations, a dashed line is drawn to connect the initial condition (particles deposited at the tumbler wall) to subsequent iterations. The mean radius drops quickly to below 0.7 in the first 20 iterations and does not vary significantly afterwards. The x-ray images at 20 iterations [Fig.~\ref{Fig:10}(b)] and 50 iterations [Fig.~\ref{Fig:10}(c)] both show x-ray opaque silver tracer particles in the middle portion of the tumbler, meaning that tracer particles relocate from near the tumbler wall to the core of the bed of particles. The dispersion of particles at 50 iterations is very similar to that at 20 iterations, and there is no apparent structure or accumulation of large particles other than near the core of the bed of particles. This result again matches the prediction of the continuum model [Fig.~\ref{Fig:4}(c)], where no significant non-mixing structures exist near the core of the particle bed. 
	
	This pair of experiments demonstrates that particles form the segregation pattern predicted by the continuum model only close to the tumbler wall. The transport of particles can also be tracked as the volume fraction of large particles increases. Similar size and density glass beads ($d$ = 3.15 mm; $\rho$ = 2.45 g cm$^{-3}$) are used along with 200 silver tracer particles ($d$ = 3.01 mm, $\rho$ = 2.5 g cm$^{-3}$) to make total large particle volume fractions $f$ = 5\%, 15\%, and 25\% mixtures with 1 mm small glass particles. Since the large glass beads and tracer particles have similar size and density, the behavior of tracer particles mirrors the bulk behavior of all large particles. The tracer particles are tracked for 50 iterations of the ($57^{\circ},57^{\circ},90^{\circ}$) protocol. The mean normalized radius of tracer particles for all volume fractions decreases only slightly with increasing large particle volume fraction, at steady state (46 - 50 iterations), as shown in Table~\ref{Tab:1}. 
	
	The result in Table~\ref{Tab:1} is consistent with the non-mixing regions expanding in size at the tumbler wall with increasing large particle volume fraction, as is evident in Fig.~\ref{Fig:5}. However, the slight reduction in the mean normalized radius in Table~\ref{Tab:1} suggests not all large particles are in the monolayer at the wall. Based on approximating the coverage of a monolayer of large particles at the tumbler wall, we estimate that most large particles are within 3 particle diameters of the wall at the higher larger particle volume fractions.  
	Thus, the large particles do not entirely occupy the conical non-mixing structures predicted by the continuum model in Fig.~\ref{Fig:4}(c) but tend to expand the regions of large particles near the tumbler wall. This indicates that the mechanism of pattern formation is closely related to the segregation dynamics that occurs in the flowing layer. The effect of segregation may be so strong that large particles are pushed to the surface of the flowing layer so they never have a chance to reach the apex of any of the conical non-mixing structures. On the other hand, when segregation is weak, particles tend to mix more uniformly instead of accumulating in the non-mixing structures as occurs for $R$  = 1.27 in Fig.~\ref{Fig:5}.

\begin{table}	
	\begin{tabular}{ c | c | c | c }
		\hline
		\hline
		 & 5\% & 15\% & 25\% \\ \hline 
		$\bar{r}$ &~ 0.9446~ &~ 0.9358 ~&~ 0.9257 ~\\ \hline
		$\sigma_r$ &~ 0.0629 &~  0.0575 & ~ 0.0816\\ \hline
		\hline		
	\end{tabular}
	\caption{\label{Tab:1} Mean normalized radius $\bar{r}$  and standard deviation of normalized radius $\sigma_r$ of tracer particles over 46 - 50 iterations of experiments with a mixture of 3 mm large silver tracer particles and glass particles and 1 mm small glass particles with  large particle volume fractions $f$ = 5\%, 15\%, and 25\%. }
\end{table}
		
		\section{Dependence on rotation Protocols}

		\begin{figure}[h]
			\includegraphics[width=0.8\columnwidth]{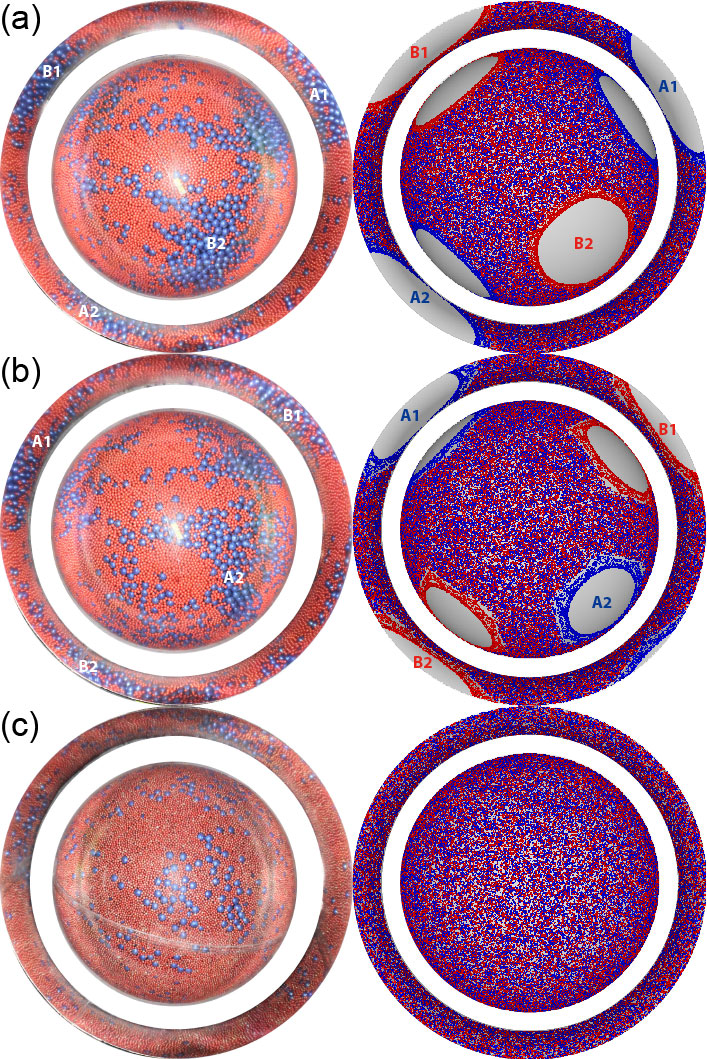}
			\caption{\label{Fig:11} Experiments (left) compared to Poincar\'e sections (right) for protocol (90$^{\circ}, 90^{\circ}, 80^{\circ}$) in (a), (90$^{\circ}, 90^{\circ}, 90^{\circ}$) in (b), and  (75$^{\circ}, 60^{\circ}, 90^{\circ}$) in (c). Visualization of tumbler wall for corresponding experiments with 15\% large blue particles (4 mm) and 90\% small red particles (1 mm).  }
		\end{figure}
		
		Similarity between the pattern formation in experiments and non-mixing regions in the continuum model can be demonstrated in many other protocols, including those shown in Fig.~\ref{Fig:11}: (90$^{\circ}, 90^{\circ}, 80^{\circ}$), (90$^{\circ}, 90^{\circ}, 90^{\circ}$), and (75$^{\circ}, 60^{\circ}, 90^{\circ}$). Large blue particles accumulate in two pairs of period-2 non-mixing islands labeled as A1-A2 and B1-B2 in Fig.~\ref{Fig:11}(a) for protocol (90$^{\circ}, 90^{\circ}, 80^{\circ}$). Note that here the two rotation axes are not orthogonal, but at an angle of $80^{\circ}$. Similarly for the orthogonal axes case (90$^{\circ}, 90^{\circ}, 90^{\circ}$), large blue particles also accumulate into period-2 non-mixing islands despite more particles appearing in the chaotic region between the non-mixing islands. For the protocol (75$^{\circ}, 60^{\circ}, 90^{\circ}$) [Fig.~\ref{Fig:11}(c)], where no non-mixing structures are predicted by the continuum model, the experiment reveals no clear segregation patterns, as expected.

		\begin{figure}
			\includegraphics[width=0.8\columnwidth]{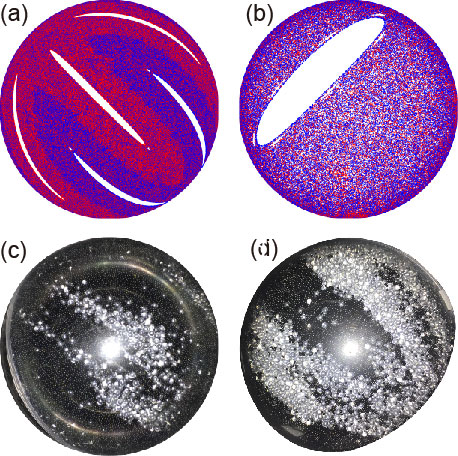}
			\caption{\label{Fig:12}Poincar\'{e} section of the continuum model in (a) the bottom view of the bulk and (b) the bottom view of the flowing layer. Segregation patterns after 30 iterations in a half-full spherical tumbler with $10\%$ large clear particles (3 mm) and $90\%$ small black particles (1 mm) starting with (c) large particles in the bottom of tumbler (bottom view), and (d) large particles on top of flowing layer (oblique view) for ($45^{\circ}, 45^{\circ}, 90^{\circ}$) protocol.}
		\end{figure}

		The protocols ($ 57^{\circ}, 57^{\circ}, 90^{\circ}$), ($ 90^{\circ}, 90^{\circ}, 80^{\circ}$), and (90$^{\circ}, 90^{\circ}, 90^{\circ}$) have what are called persistent non-mixing islands \cite{Zaman2018}. They correspond to cells in PWI maps \cite{Smith2017,Park2016}, which provide a mathematical description of the action of cutting-and-shuffling that forms the foundation for non-mixing regions in spherical tumbler flow with alternating rotations about two axes \cite{Zaman2018}. Although PWI maps assume a non-physical infinitely thin flowing layer at the surface, the structures can persist for a finite-thickness flowing layer, as is evident in these experiments. The boundaries of the non-mixing islands are defined by the interface between the flowing layer and the bulk \cite{Zaman2018,Smith2017}. 
		
		In addition to the persistent non-mixing regions described thus far, there is another type of barrier to mixing known as an emergent non-mixing region \cite{Zaman2018}. Unlike persistent non-mixing regions that pass entirely through the flowing layer with each rotation and have boundaries set by the boundary of the flowing layer, emergent non-mixing regions periodically land entirely within the flowing layer, are stretched in the streamwise direction, and have boundaries that are not necessarily coincident with the flowing layer boundary \cite{Zaman2018}.
	    For a detailed discussion on persistent and emergent non-mixing islands, see Zaman et al.~\cite{Zaman2018}. An example of a Poincar\'{e} section exhibiting emergent non-mixing islands is shown in Fig.~\ref{Fig:12}(a,b) for protocol ($45^{\circ}, 45^{\circ}, 90^{\circ}$). The Poincar\'{e} section consists of interpenetrating fingers that are primarily red or primarily blue when viewed from the bottom [Fig.~\ref{Fig:12}(a)]. The boundaries between the red and blue regions are shown to be barriers to mixing \cite{Zaman2018}. The stretching of one of the white non-mixing regions at the centers of the interpenetrating fingers is evident in Fig.\ref{Fig:12}(b), which shows a non-mixing region as visualized from below as it lands entirely in the flowing layer between rotations.

		Segregation experiments with a mixture of $10\%$ large clear glass particles ($d$ = 2.97 $\pm$ 0.05 mm) with $90\%$ small black glass particles ($d$ = 1.05 $\pm$ 0.05 mm) are carried out for the ($45^{\circ}, 45^{\circ}, 90^{\circ}$) protocol with two different initial conditions: large particles initially located in the bottom of the tumbler on the tumbler wall  or large particles initially spread on top of the flowing layer. In the first experiment, large particles initially located at the bottom on the tumbler wall, accumulate into three pairs of thin strips, with the most visible pair in the middle [Fig.~\ref{Fig:12}(c)]. This pattern corresponds to the red fingers in the continuum model [Fig.~\ref{Fig:12}(a)] with a thin non-mixing region in the middle of the fingers. In Fig.~\ref{Fig:12}(c) the large particles form two thin strips as they accumulate in the middle red finger in Fig.~\ref{Fig:12}(a). The white non-mixing region in the middle of the finger [Fig.~\ref{Fig:12}(a)] is reflected in Fig.~\ref{Fig:12}(c) as the central small black particle filled region.  In the second experiment [Fig.~\ref{Fig:12}(d)], large particles initially located in the flowing layer accumulate in the two visible blue fingers of the Poincar\'{e} map, again with the central white region filled with small black particles [a third blue finger is stretched across the flowing layer, consistent with  Fig.~\ref{Fig:12}(b)]. This result is expected since the blue finger is mapped onto the flowing layer where large particles initially are located [Fig.~\ref{Fig:12}(a)]. In both experiments, large particles do not occupy the white non-mixing regions predicted in the continuum model. Instead they accumulate in the colored finger like structures, demonstrating that the mixing barriers between the red and blue dominant regions in the continuum model are physical mixing barriers preventing material exchange \cite{Zaman2018}. Particles starting in one region are mostly confined in this region even with collisional diffusion.
		For all of the protocols in this section, the experiments demonstrate robust pattern formation matching features in the Poincar\'{e} sections derived from the continuum model.

	\section{Mechanism of pattern formation}	
	\subsection{Pattern formation in 3D versus quasi-2D}
	At first glance, segregation patterns in quasi-2D tumbler and 3D spherical tumbler appear similar. In the flowing layer, particles segregate due to differences in size, and at steady state one species accumulates in the predicted non-mixing islands while the other species occupies the rest of the domain. However, a major difference is that large particles occupy the chaotic region in a quasi-2D tumbler, while they accumulate in non-mixing structures in a 3D spherical tumbler for persistent non-mixing structures. Conversely, small particles accumulate in the non-mixing region in a quasi-2D tumbler, while they occupy the chaotic region in a 3D spherical tumbler. 
	
	\begin{figure}[h]
		\includegraphics[width=0.8\columnwidth]{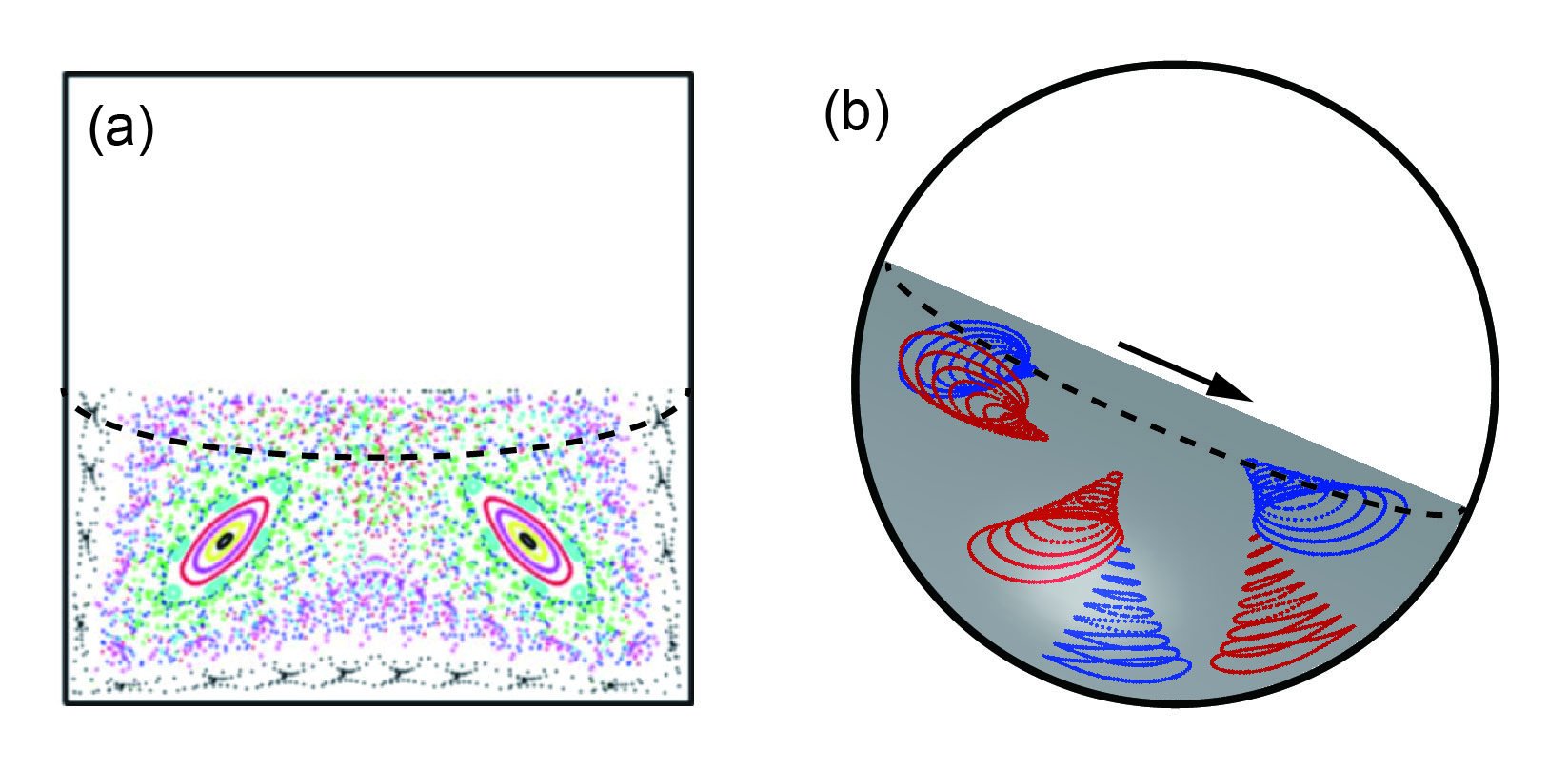}
		
		\caption{\label{Fig:13}Schematic of the boundary of the flowing layer (dashed curve) superimposed onto the Poincar\'{e} section for (a) a half-full quasi-2D square tumbler and (b) a half-full 3D spherical tumbler.}
	\end{figure}
	
	To explain this difference, first note that segregation in the flowing layer occurs in the same manner in both quasi-2D and 3D spherical tumbler geometries. Small particles percolate down to the bottom of the flowing layer, while large particles rise to the top. The small particles deposit on the static bed of particles first, while the top layer of large particles continues to flow down the free surface. As a result, small particles are deposited near the  middle of the flowing layer, while large particles are deposited further downstream. This deposition pattern is reinforced through periodicity due to the tumbler shape (2D) or due to the biaxial protocol(3D). The key point is that the bottom of the flowing layer has mostly small particles, which in the quasi-2D tumbler is where the non-mixing islands overlap the flowing layer as shown schematically in  Fig.~\ref{Fig:13}(a). Consequently, small particles fall into the non-mixing islands when they segregate to the bottom of the flowing layer. Even if they segregate to the bottom of the flowing layer, where a non-mixing island is not present, they continue to be advected throughout the chaotic flow region (or via collisional diffusion), and eventually make their way into the non-mixing islands in subsequent rotations. 
	
	In contrast, the conical non-mixing structures in a 3D spherical tumbler have significant volume near the tumbler wall [Fig.~\ref{Fig:13}(b)]. This means that the conical base of the non-mixing regions corresponds to where large particles tend to segregate. As a result, the large particles tend to fill the non-mixing structures in the 3D spherical tumbler. Thus, how the particle segregation in the flowing layer coincides with the non-mixing islands determines which particles accumulate in non-mixing regions. That is, the interplay between underlying advection field and the particle segregation in the flowing layer determines the ultimate segregation pattern. Whether it is large particles or small particles accumulating in the non-mixing regions depends on the relative location of non-mixing regions when they pass through the flowing layer. 
	
	Another important point is that the accumulation of particles in non-mixing regions is made possible when the radial segregation is in the same plane as the mixing barriers. The segregation-driven material exchange across mixing barriers into the non-mixing regions inherently occurs in a single plane in a quasi-2D tumbler. However, the 3D spherical tumbler is more complicated because of its 3D nature. Along with the radial transport, the large particles that accumulate into the non-mixing regions must also align with the non-mixing regions in the spanwise direction, requiring axial displacement of large particles.

\begin{figure}[ht]
	 	\includegraphics[width=0.5\columnwidth]{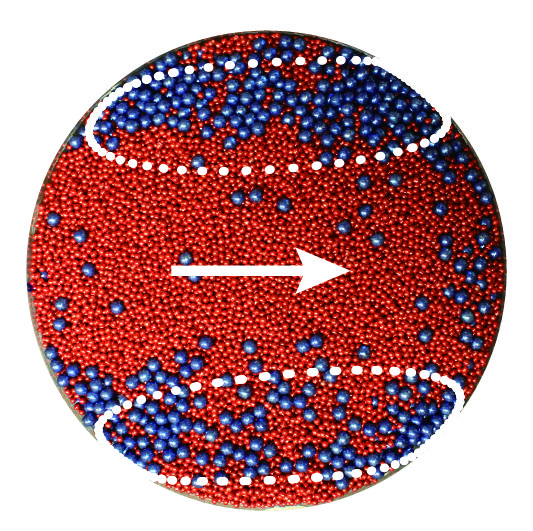}
	 	\caption{\label{Fig:14} Top view of the flowing layer in a tumbler half-filled with 15\% large blue particles (4 mm) and 85\% small red particles (2 mm) after 20 revolutions about a single axis. White dotted curves represent boundaries of two non-mixing islands predicted for the protocol (57$^{\circ}, 57^{\circ}, 90^{\circ}$) as they would appear when flowing across the free surface predicted by the continuum model. The white arrow indicates the direction of the flow.}
	 \end{figure}

	In previous studies of single axis rotation of a spherical tumbler partially filled with size bidisperse mixtures \cite{Naji2009,Chen2009,DOrtona2016a}, two different axial segregation patterns occur for a mixed initial condition depending on the fill fraction, absolute particle sizes, particle size ratio, volume fraction of large particles, or smoothness of tumbler wall. Large particles either accumulate in a band at the equator of the tumbler, or in two bands near the poles. For the operating condition of this study (rotation speed 2.6 rpm and half-filled tumbler), large particles tend to accumulate at the poles when the tumbler is continuous rotated about a single axis. This is shown in a top view of the free surface after 20 revolutions in Fig.~\ref{Fig:14}. The large particles experience a small axial drift velocity that gradually drives them to concentrate in the bands near the poles, consistent with previous experimental results \cite{Chen2009,DOrtona2016a}. 
	 
	 Non-mixing islands predicted by the continuum model for the biaxial (two-axis)  rotation protocol as they pass through the flowing layer are shown by the white dotted boundaries overlaid on the single axis  experiment result in Fig.~\ref{Fig:14}. The axial positions of the non-mixing islands coincide with the regions of large particle accumulation for single axis rotation. 
	 Of course, the bands formed for the single axis rotation experiment is a steady state result after 20 tumbler revolutions. In the biaxial experiments that we focus on in this paper, each single axis rotation action is smaller than $90^{\circ}$. Thus, the axial segregation is small for each iteration, yet the axial segregation is reinforced through repeated iterations. For each single axis action about either the $z$-axis or $x$-axis, the large particles segregate toward the poles and accumulate in the coincident non-mixing islands. Segregation in both the radial and spanwise directions ensures that large particles move to the free surface and toward the poles where non-mixing structures have the largest volume. When this concentrating effect due to segregation is stronger than collisional diffusion, such as in the cases of large size ratios, the segregation pattern is clear and the features have distinct boundaries as shown for large size ratios in Fig.~\ref{Fig:5}. Similar to radial segregation of small particles in the quasi-2D square tumbler, both radial and axial segregation aligns the large particles with non-mixing regions in the 3D spherical tumbler. Large particles are concentrated in these regions with repeated rotations. It is precisely the alignment of segregation with non-mixing regions that allows the large particles to accumulate in non-mixing regions.

	\subsection{\label{Sec:D}A dynamical systems perspective}
	
	\begin{figure*}
		\includegraphics[width=0.8\textwidth]{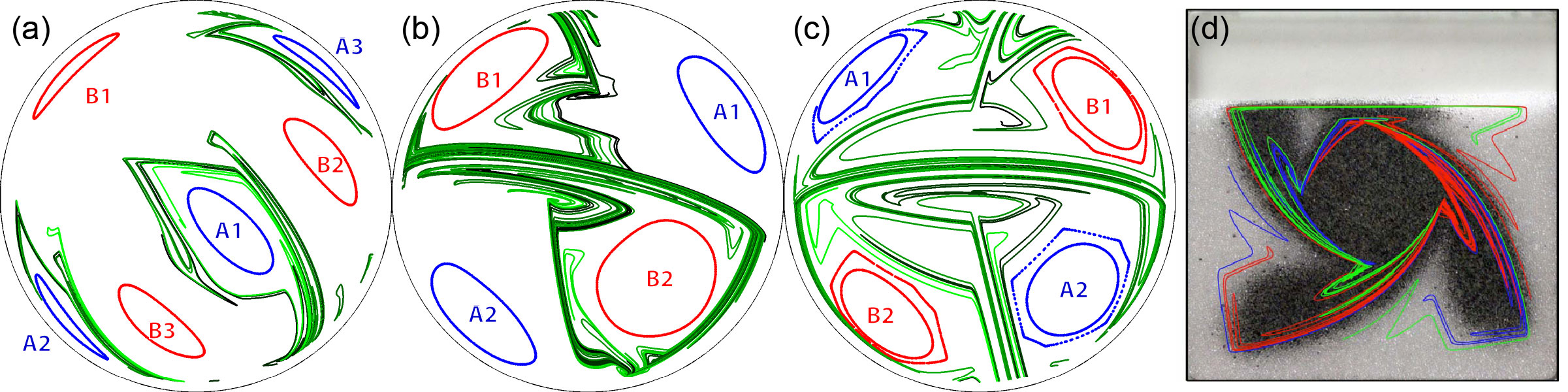}
		\caption{\label{Fig:15}Unstable manifolds of period 3 in gradient color from black to green for protocol (a) (57$^{\circ}, 57^{\circ}, 90^{\circ}$), (b)(90$^{\circ}, 90^{\circ}, 80^{\circ}$) in (b), and (c) (90$^{\circ}, 90^{\circ}, 90^{\circ}$). The blue and red closed curves are boundaries of non-mixing islands. Both the unstable manifolds and non-mixing islands are generated on the $r$  = 0.95 surface. (d) Unstable manifolds superimposed onto experiment of a 75\%-full quasi-2D tumbler with a mixture of 30\% small black particles. Reprinted with permission from Meier et al. \cite{Meier2006} \copyright 2018 American Physical Society.}
	\end{figure*}

	Experiments for protocols (57$^{\circ}, 57^{\circ}, 90^{\circ}$), (90$^{\circ}, 90^{\circ}, 80^{\circ}$), and (90$^{\circ}, 90^{\circ}, 90^{\circ}$) demonstrate robust pattern formation near the tumbler wall [see Fig.~\ref{Fig:4}, \ref{Fig:5}, and \ref{Fig:11}(a-b)]. Particle accumulation for protocol (57$^{\circ}, 57^{\circ}, 90^{\circ}$) is the most compact and has the most well defined boundaries despite the fact that the non-mixing islands are the smallest in size predicted by the continuum model. On the other hand, the patterns for protocols (90$^{\circ}, 90^{\circ}, 80^{\circ}$) and particularly protocol (90$^{\circ}, 90^{\circ}, 90^{\circ}$) have less clearly defined boundaries [Fig.~\ref{Fig:11}(a-b)], even though the islands predicted by the continuum model are larger. This result occurs regardless of the large particle fraction, so it cannot be explained on that basis. To explore this contradiction, we examine flow behavior in the chaotic region immediately surrounding the non-mixing islands, neglecting effects of axial segregation and diffusion because they are small compared to the mean flow. The chaotic behaviors are induced by complex structures formed from stable and unstable manifolds, which correspond to contraction and expansion of material around a hyperbolic fixed point \cite{Ottino1989}. The algorithm used to compute the manifolds is described in Appendix \ref{UM}.
	
	The unstable manifolds calculated from the continuum model for the portion of the tumbler visible from the bottom are shown with the non-mixing islands for protocols (57$^{\circ}, 57 ^{\circ}, 90^{\circ}$), (90$^{\circ}, 90^{\circ}, 80^{\circ}$), and (90$^{\circ}, 90^{\circ}, 90^{\circ}$) in Fig.~\ref{Fig:15}. There are two unstable manifolds for each of the three protocols, which correspond to the two sets of periodic points. For simplicity, only one of the two manifolds is shown in Fig.~\ref{Fig:15}. The separate portions of the manifold shown in each of the images are connected through the flowing layer, which is not shown. As the unstable manifolds stretch from the hyperbolic points, the color progresses from black to green. 
	One unstable manifold of protocol (57$^{\circ}, 57^{\circ}, 90^{\circ}$) [Fig.~\ref{Fig:15}(a)] wraps tightly around one group of the non-mixing islands (A1-A2-A3). Particles that happen to be bumped out of the non-mixing regions by collisional diffusion are advected by the unstable manifold. In subsequent iterations, the chance of the particle diffusing (or being advected) back into the non-mixing regions are high because the manifold wraps tightly around the A1-A2-A3 non-mixing regions. For protocol (90$^{\circ}, 90^{\circ}, 80^{\circ}$), even though the unstable manifold wraps around the non-mixing islands B1-B2 [Fig.~\ref{Fig:15}(b)], the unstable manifold linkage between the two non-mixing islands occupies a larger portion of the domain, spanning the entire tumbler horizontally. Therefore, the accumulation of large particles for protocol (90$^{\circ}, 90^{\circ}, 80^{\circ}$) is less distinct [Fig.~\ref{Fig:11}(a)] than for protocol (57$^{\circ}, 57^{\circ}, 90^{\circ}$) [Fig.~\ref{Fig:4}(a)] because a particle that starts outside of a non-mixing region or is bumped outside of the non-mixing region by collisional diffusion is more likely to be dispersed by the manifold to positions elsewhere in the chaotic region. For protocol (90$^{\circ}, 90^{\circ}, 90^{\circ}$) [Fig.~\ref{Fig:15}(c)], the unstable manifold spans the entire domain and forms large folds that encompass all four non-mixing islands. Thus, particles are more likely to be dispersed to regions elsewhere in the chaotic sea once they are carried away by the manifolds, as is evident in the less distinctly segregated regions for the experiment of protocol (90$^{\circ}, 90^{\circ}, 90^{\circ}$) [Fig.~\ref{Fig:11}(b)].

	The relation between  unstable manifolds and the segregation pattern is also evident in the quasi-2D tumbler. As shown in Fig.~\ref{Fig:15}(d), the unstable manifolds surrounding the non-mixing islands outline the boundary of the segregation pattern. The unstable manifolds serve as a mixing barrier in that particles follow the manifolds instead of moving across them. This effect is observed in the 3D spherical tumbler for protocol (57$^{\circ}, 57^{\circ},90^{\circ}$). In contrast, for protocol (90$^{\circ}, 90^{\circ}, 90^{\circ}$), the unstable manifolds fill the entire domain, carrying particles throughout  rather than acting as a barrier to mixing. As a result, the unstable manifolds in this case facilitate material transport throughout the domain, resulting in an indistinct segregation pattern [Fig.~\ref{Fig:11}(b)]. 
	
	\section{Discussion \& Conclusions}
	
	We started this paper by asking if granular segregation and chaotic dynamics can interact in a fully 3D system to generate segregation patterns, and the answer is clearly ``yes'' based on the visualization experiments alone. To further understand the relation between the segregation and chaotic dynamics, the segregation patterns of granular flows in a 3D spherical tumbler are examined with both surface visualization and x-ray imaging. Both methods demonstrate particles of different sizes segregate into patterns predicted by a simple kinematic continuum model independent of any segregation model. The detailed characteristics of the segregation pattern depend on relative strengths of segregation, diffusion, and chaotic advection, as is evident from experiments with different particle size ratios, mixture fractions, and rotation protocols. 
	
	The pattern formation is a result of the coincidence between the non-mixing islands of the underlying flow field and particle segregation for both the quasi-2D square tumbler studied previously \cite{Meier2006} and the 3D spherical tumbler examined here.  When the direction of accumulation due to segregation aligns with non-mixing structures of the underlying flow field, segregation patterns form. 
	
	Other factors can influence the details of the pattern, including the strength of the segregation, collisional diffusion, and unstable manifold transport. These effects can be tuned to change the pattern to various degrees. For example, the relative strength of segregation compared to diffusion can be manipulated in several ways including changing rotation speed, particle to tumbler size ratio, and relative particle sizes. For instance, the pattern becomes weak when segregation is decreased with a smaller particle size ratio and thus a higher relative influence of collisional diffusion. It is also likely that when axial segregation is weaker, the manifold transport becomes more important. The segregation pattern is perturbed more easily for protocol (90$^{\circ}, 90^{\circ}, 90^{\circ}$) than protocol (57$^{\circ}, 57^{\circ}, 90^{\circ}$), because the unstable manifolds are capable of carrying particles further away from the non-mixing regions. Hence, segregation pattern formation depends not only on  the structure of dynamical systems features including elliptic non-mixing regions and unstable manifolds, but also on the relative strength of segregation compared to collisional diffusion. The relative strengths of these factors is key to predicting the pattern formation or to optimizing the mixing by avoiding such pattern formation.
	
	It is important to realize that 3D flow in this study generates 2D invariant structures on hemispherical shells. Therefore, radial transport occurs only via diffusion and segregation.  Fully 3D transport including radial transport can occur via streamline jumping \cite{Christov2010,Christov2014}, which can be introduced to the experiments, for example, by rotating the tumbler about two axes at different speeds. Thus, 3D spherical tumbler flow can serve as a prototypical system for studying pattern formation and mixing in even more complicated 3D dynamical systems.

%	\nocite{apsrev41control} 

\begin{acknowledgments}
	M. Y. thanks Lachlan D. Smith for valuable discussions. This research was funded by NSF Grant No. CMMI-1435065.
\end{acknowledgments}

\appendix
\section{\label{CM}The continuum model}
In the continuum model \cite{Christov2014,Zaman2018}, particles flow down the surface in a thin flowing layer that lies on top of non-flowing particles in the bulk that move in solid body rotation with the tumbler. The flow is simplified to two-dimensional in the streamwise direction. The flowing layer velocity is approximated by a constant shear rate velocity profile. For rotation about any axis (the $z$-axis here with $x$ in the streamwise direction and $y$ normal to the free surface), the non-dimensionalized velocity field  \textbf{u} = ($u,v,w$) is piecewise defined such that the flowing layer (0$\ge$y$\ge$-$\delta$) velocity is \textbf{u}\textsubscript{fl} = (($\delta$ + y)/$\epsilon$$^{2}$, $xy/\delta$, 0) and the bulk (y $< -\delta$) solid body rotation velocity is \textbf{u}\textsubscript{b} = ($y,-x, 0$). The interface of the lenticular flowing layer with the bulk is located at $\delta$($x,z$) = $\epsilon\sqrt{1-x^2-z^2}$, where $\epsilon = \delta(0,0) = \sqrt{\omega/\dot{\gamma}}$ is the maximal dimensionless flowing layer depth at the center of the sphere ($x=z=0$) for shear rate $\dot{\gamma}$ and angular rotation velocity $\omega$. All variables are dimensionless---lengths are normalized by the tumbler radius $R_0$ and the rotation period $T$ is normalized by 1/$\omega$. Note that we assume the particles gain the velocity described above immediately as the tumbler starts to rotate, and stop moving instantaneously as the tumbler stops rotating. This continuum model, which includes stretching characteristic of chaotic flows \cite{Ottino1989}, is parameterized by the flowing layer depth $\epsilon$, which is set to 0.15 to match the conditions in the experiments ($\omega$ = 2.6 rpm). Experiments in this work are done with the same rotation speed about both $x$-axis and $z$-axis, so that $\epsilon_x = \epsilon_z$. 
\section{\label{PS}Poincar\'{e} section}
Stroboscopic maps of 500 iterations are used to investigate mixing and non-mixing behaviors of the underlying flow. Positions of initial points are recorded after every iteration of biaxial rotations calculated by the continuum model. 
The initial conditions are points seeded on the interface between the static bed and the flowing layer before the 1st $z$-axis rotation in blue and before the first $x$-axis rotations in red on the same radial hemispherical shell. Piecewise analytical solutions for the differential velocity profile are derived in \cite{Christov2014} for the bulk and flowing layer. The chaotic region manifests itself as points advected throughout the domain, e.g.  Fig.~\ref{Fig:3}(b), while the white elliptical regions avoided by the tracer points represent elliptic non-mixing regions that prevent material exchange. In the 3D spherical tumbler rotated with the same speed about both $z$-axis and $x$-axis, the Poincar\'{e} section exists on invariant surfaces parametrized by the radius of hemispherical shells \cite{Christov2014}. In other words, tracer points have trajectories that lie on the same radial surfaces they start on.  

\section{\label{UM}Unstable manifolds}
The unstable manifolds are traced by tracking points seeded on a short line segment of length 0.001$R_{o}$ in the direction corresponding to material expansion of the hyperbolic point. Positions of tracer points are recorded after every iteration, and the resulting manifolds shown for each protocol in Fig.~\ref{Fig:15} are trajectories of tracer points advected for 15 iterations. In order to maintain a uniform density tracing in the presence of fast manifold stretching, new points are back inserted in intervals between consecutive points that are $5\times 10^{-5}$$R_0$ apart or further after each iteration. 

The eigenvalues and eigenvectors of the hyperbolic points are first calculated from the Jacobian of the mapping $\Phi^n (x) = x$ of the biaxial tumbler flow \cite{Christov2014,Smith2016-2}. For the volume-preserving map studied here, the three eigenvalues have a product of $\lambda_1\lambda_2\lambda_3 = 1$. There is a null direction at each periodic point providing a local invariant, corresponding to the eigenvalue of 1, $\lambda_1 = 1$ \cite{Smith2016-2}. A hyperbolic point also has two real eigenvalues whose product is 1, $\lambda_2 = 1/\lambda_3$. Material expands along the direction corresponding to eigenvalue $\lambda >$ 1, and contracts along the direction corresponding to eigenvalue $\lambda <$ 1 \cite{Moser1973}. The unstable manifold associated with a hyperbolic point consists of all points that converge to the hyperbolic point as the number of iterations of the map approaches negative infinity.

\bibliography{library}

\end{document}